\documentclass[aps,pre,preprint,showpacs,floatfix]{revtex4}
\usepackage{graphicx,bm,amsmath,dcolumn,}
\begin{document}
\newcommand{\half}{\frac{1}{2}}
\newcommand{\mat}[1]{{\bf #1}}
\newcommand{\vect}[1]{\stackrel{\rightarrow}{#1}}
\newcommand{\thrd}{\frac{1}{3}}
\newcommand{\twth}{\frac{2}{3}}
\newcommand{\gae}{ {}^>_{\sim}}
\newcommand{\lae}{ {}^<_{\sim}}
\title{Crossing bonds in the random-cluster model}
\author{
Wenan Guo~$^{1}$ 
Youjin Deng~$^{2}$ 
Henk W.J. Bl\"ote~$^{3}$
}
\affiliation{$^{1}$Physics Department, Beijing Normal University,
Beijing 100875, P. R. China}
\affiliation{$^{2}$Hefei National Laboratory for Physical
Sciences at Microscale,
Department of Modern Physics, University of Science and
Technology of China, Hefei 230027, China }
\affiliation{ $^{3}$Lorentz Institute, Leiden University,
  P. O. Box 9506, 2300 RA Leiden, The Netherlands}             
\date{\today} 
\begin{abstract}
We derive the scaling dimension associated with crossing bonds in the
random-cluster representation of the two-dimensional Potts model,
by means of a mapping on the Coulomb gas. The scaling field associated
with crossing bonds appears to be irrelevant, on the critical as 
well as on the tricritical branch. The latter result stands in a
remarkable contrast with the existing result for the tricritical O($n$)
model that crossing bonds are relevant. In order to obtain independent 
confirmation of the Coulomb gas result for the crossing-bond exponent,
we perform a finite-size-scaling analysis based on numerical
transfer-matrix calculations.
\end{abstract}
\maketitle 

\section {Introduction }\label{intro}
In general one may expect that there exist large regions in the
parameter space of the Hamiltonian of a critical system that belong
to a single universality class. Or, stated in a different way, the
critical exponents do, in general, not depend on the microscopic
details of the Hamiltonian. The microscopic details will usually
only contribute to the irrelevant fields as defined in the
renormalization theory \cite{Wilson}, and thereby influence the
correction-to-scaling amplitudes.
It is therefore surprising that it was found that the introduction of
next-nearest-neighbor interactions in certain two-dimensional O($n$)
models, specified below, does affect the {\em leading} critical behavior.

In most exactly solved O($n$) loop models, the loops do not cross
or intersect, so that the loops are not entangled.
The introduction of next-nearest-neighbor interactions, i.e.,
crossing bonds, in the square-lattice O($n$) model leads, however,
to a different situation. The crossing of two different loops
segments produces a `defect' that can be neutralized at a second
crossing of the loop segments, or at an ordinary vertex where two
of the loop segments emerging from the intersection point connect.
Here the word `intersection' refers to the projection of the loop
segments on a planar lattice, the two loop segments should be
considered as to remain separated from one another.

In the analysis of connected correlation functions between two
crossings, one has to treat these loop crossings as topological
defects, i.e., the annihilation of a defect at an ordinary vertex
has to be disabled. Thus, we have a vertex with four outgoing loop
segments at position $0$, and the four segments come in at a vertex
at position $r$. It is the `watermelon' diagram with four lines.

After a mapping on the Coulomb gas, these two loop crossings are
represented by known electric and magnetic charges \cite{BN2,SD},
so that the renormalization exponent associated with the
fugacity of the crossings follows immediately. It is the same
exponent as the one describing cubic crossover in the O($n$) model,
$y=2-2g+(1-g)^2/(2g)$, where $g$ is related to $n$ by 
$n=-2 \cos(\pi g)$, with $1<g<2$ for the critical branch, and with 
$0<g<1$ for the low-temperature branch.
It is irrelevant in the critical two-dimensional O($n$) model with
$n<2$, but it becomes marginal at $n=2$, and relevant in the 
low-temperature branch of the O($n$) model \cite{BN2}. It was found by 
Jacobsen et al.~\cite{JRS} that crossing bonds do indeed induce crossover
to a new universality class \cite{MNR} in the low-temperature branch.
Another recent result establishes a relation between the partition
sum of the low-temperature branch of the O($n$) model and that of a
tricritical O($n$) model \cite{NGB}. The exponent associated with 
crossing bonds remains unchanged under this mapping. Therefore,
crossing bonds are relevant in the tricritical O($n$) model.
However, the case $n=1$, i.e. the tricritical Ising model, is not
believed to be unstable with respect to crossing bonds, and is
therefore interpreted as a special case, apparently because the
truncation of the spin dimensionality to $n=1$ in effect eliminates
the amplitudes associated with the effects of scaling fields acting
on the other spin components.

These findings for the low-temperature and the tricritical O($n$)
model raise the question what are the consequences of crossing
bonds in the two-dimensional Potts model \cite{RBP,FYW} and the
equivalent random-cluster model \cite{PWK}, and also in the related
tricritical models. The Potts model is defined in terms of lattice
variables $\sigma_i$ that can assume the discrete values 1, 2,
$\cdots$, $q$, and the index $i$ refers to the lattice site.
The reduced Hamiltonian of the model is
\begin{equation}
{\mathcal H}/k_{\rm B}T = -  K \sum_{\langle{ij}\rangle}
\delta_{{\sigma}_{i}, {\sigma}_{j}},
\label{Potts}
\end{equation}
where $K$ is the Potts coupling, inversely proportional to the
temperature. The summation indicated by ${\langle{ij}\rangle}$ is
over all interacting nearest-neighbor pairs of  Potts variables.
A generalization to continuous $q$ is obtained by mapping this
model (\ref{Potts}) onto the Kasteleyn-Fortuin random-cluster
model \cite{PWK}, whose partition sum is
\begin{equation}
Z(q,K)=\sum_{\{b\}} u^{n_b} q^{n_c},
\label{Zrc}
\end{equation}
where the sum is over all graphs $\{b\}$ formed by independent bond
variables (absent or present), the bond weight is given by
$u=1-e^{-K}$, $n_b$ is the number of present bonds, and $n_c$ is the
number of clusters formed by these bonds.
The number $q$ of Potts states now appears as a continuous variable,
so that the random-cluster model can be seen as a generalization of the
Potts model. In the limit $q\to 1$ it reduces to the bond-percolation model.
The presence of crossing bonds in a percolation model \cite{FDB} has
been investigated, but that work did not yield evidence for a possible
modification of the finite-size-scaling behavior.

This question about the effects of crossing bonds in this model
will be answered below. In Sec.~\ref{CG}
we derive the scaling dimension associated with crossing bonds
analytically, with the help of a renormalization mapping of the Potts
model on the Coulomb gas. The validity of this analytic result is
checked numerically in Sec.~\ref{TM}, by means of a transfer-matrix
analysis combined with finite-size scaling. A short discussion in
Sec.~\ref{Dis} concludes this paper.

\section {Coulomb gas }\label{CG}
\begin{figure}
\includegraphics[scale=0.5]{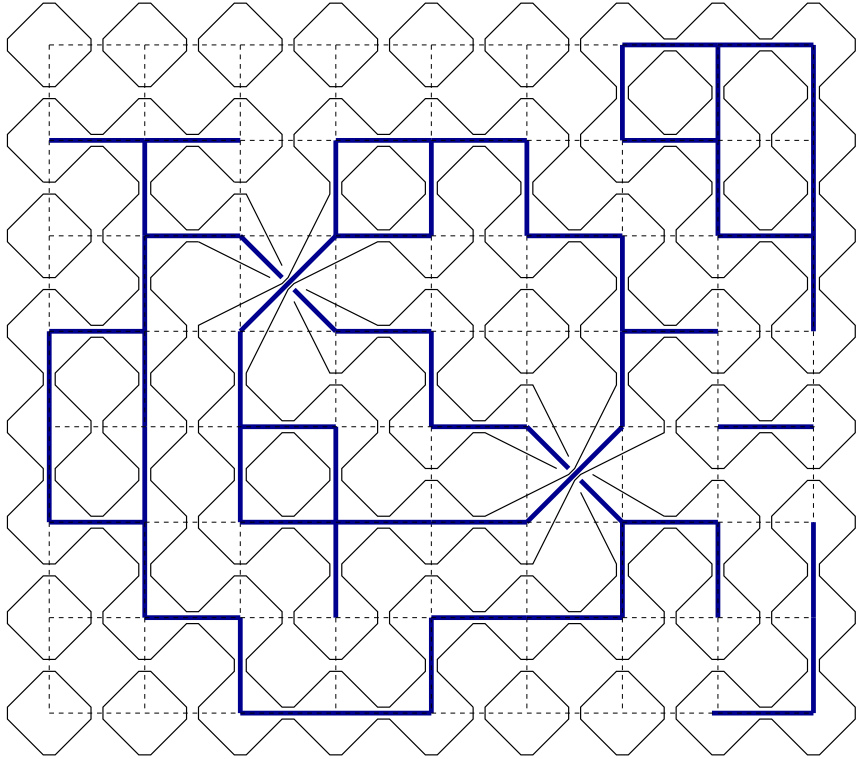}
\caption{Random-cluster configuration with two crossing-bond vertices,
and the corresponding loop configuration on the surrounding lattice.
The two vertices are connected by four random clusters, each of which
is surrounded by two loop segments. The two vertices are thus connected
by eight loop segments.}
\label{figc}
\end{figure}

The mapping of the random-cluster model on the Coulomb gas involves,
as a first step, the representation of the random clusters by means of
loops on the surrounding lattice. For a random-cluster model without
crossing bonds, one obtains a system of non-intersecting loops \cite{BKW}. 
Let us now consider the `defect' introduced by the crossing or intersection
of two random-cluster bonds. The four random-cluster bond segments going
out from the intersection point are represented by {\rm eight} outgoing 
loop segments. This is an important difference with the O($n$) loop model,
where one has four outgoing loop segments. While these crossing bonds 
introduce an entangled random-cluster configuration, the entanglement
can be eliminated at ordinary vertices where random-cluster bonds
meet. However, we are interested in the connected correlation function
$g(r)$ associated with two of these crossing-bond vertices separated
by a distance $r$. Thus, we have to treat the defects as topological
defects, and disable the annihilation of entanglement at ordinary
vertices. The two special vertices are connected by four different
random clusters. In the language of the surrounding loop model, 
we have to analyze the watermelon diagram with eight legs, as 
illustrated in Fig.~\ref{figc}.

In the Coulomb gas, both of these vertices are represented by a 
charge consisting of an electric and a magnetic component \cite{BN2}. 
At position 0 and $r$, these charges are denoted $(e_0,m_0)$ and
$(e_r,m_r)$ respectively. Here, $m_0=-m_r=4$ is half the number of loop
segments of the watermelon diagram, and $e_0=e_r=1-g$ is determined by
the Coulomb gas coupling constant $g$ which is known as a function
of the number of Potts states of the critical random-cluster model:
\begin{equation}
g=1-\frac{1}{2 \pi} {\rm arccos}\left( \frac{q}{2}-1\right) \, .
\label{gdef}
\end{equation}
The decay of the correlation function $g(r)$ as a function of $r$ is
as $g(r) \propto r^{-2X_x}$, governed by the scaling dimension $X_x$.
The scaling dimensions associated with a general pair of charges 
is given by 
\begin{equation}
X(e,m) = - \, \frac{e_0 e_r}{2 g} - \frac{m_0 m_r g}{2}
\label{Xem}
\end{equation}
which implies, in the present case,
\begin{equation}
X_x=1-\frac{1}{2g}+\frac{15 g}{2} \, .
\label{Xx}
\end{equation}

\section {Transfer-matrix analysis }\label{TM}
Transfer-matrix calculations usually apply to a system wrapped on
a cylinder with a finite circumference $L$, while the limit is taken
of an infinite length.
In transfer matrix calculations on random-cluster models, the concept
of `connectivity' is an essential ingredient. For a square-lattice
model with only nearest-neighbor couplings, wrapped on a cylinder 
with an open end, there are $L$ sites located at this end. Sites that
belong to the same cluster are said to be connected. Each of the
$L$ end sites can thus be connected to zero or more other end sites;
the precise way in which the end sites are connected is called 
`connectivity'. These connectivities can be coded by means of 
subsequent integers, which then serve as a transfer-matrix index.
We denote the partition sum of a model with $L \times M$ sites on a
cylinder as $Z^{(M)}$. Then, $Z^{(M)}$  can be divided into a number
of restricted sums according to the index $\alpha$ of the connectivity 
of the $M$th row, i.e., $Z^{(M)}=\sum_{\alpha} Z^{(M)}_{\alpha}$.
Then, the restricted sums for a model with $M+1$ layers of $L$ sites
can be written as a linear combination of the restricted sums for a 
system with $M$ layers \cite{SML}:
\begin{equation}
Z^{(M+1)}_{\alpha} = \sum_\beta T_{\alpha \beta} Z^{(M)}_{\beta}
\label{tdef}
\end{equation}
where the coefficients $T_{\alpha \beta}$ define the transfer matrix
$\mat{T}$.

The number of connectivities for a finite size $L$, as well as the
way they are coded, still depends on the type of random-cluster
model. In the absence of a magnetic field, and for nearest-neighbor
bonds only,  the connectivities are `well nested'. This means that,
if the end sites numbered $i$ and $j$ are connected, and $k$
and $l$ are connected, while the pair $i,j$ is not connected to $k,l$,
that the case $i<k<j<l$ is excluded. The random-cluster bonds are not
entangled. More details, including a full description of the coding
algorithm, appear in Ref.~\onlinecite{BN1982}.

\subsection{Entangled random clusters}
The problem defined in Sec.~\ref{intro} does, however require the
introduction of entangled random-cluster configurations. For the
general problem of a random-cluster model with crossing bonds, one has
to provide a different coding algorithm which allows only the analysis 
of a very limited range of finite sizes \cite{FDB}, because of the rapid
increase with $L$ of the required number of connectivities. A better
approach is the use of the `magnetic' connectivities, also defined in
Ref.~\onlinecite{BN1982}, which enable the introduction of a nonzero
magnetic field. While we are here not interested in the introduction
of a field, we do use the property of these magnetic connectivities
that they allow for one magnetic cluster that need not
be well nested in combination with other clusters. The occurrence 
of one such a cluster is sufficient for our present purposes.
The number of these magnetic connectivities increases less rapidly
with $L$ than that of the entangled connectivities
used in Ref.~\onlinecite{FDB} for a crossing-bond model.

In our transfer-matrix analysis, we have to perform the calculation
of the exponent associated with the connected correlation function
between two vertices with four outgoing random-cluster bonds. 
Thus there exist four random clusters, which are not mutually
connected (except at the two vertices), and each of these clusters
connects to precisely one outgoing bond of each of the two vertices.
The calculation of this exponent can be done by means of transfer
matrices, with the help of Cardy's conformal mapping \cite{Cardy-xi}
between the infinite plane and a cylinder with circumference $L$.
The mapping of a conformally invariant model between these two
geometries shows that the scaling dimension $X_x$ describing the
algebraic decay of the correlation function $g(r)\propto r^{-2X_x}$
in the infinite plane is related to the length scale $\xi_x(L)$
describing the exponential decay of the analogous correlation
function with distance in the cylindrical geometry \cite{Cardy-xi} as
\begin{equation}
X_x \simeq  \frac{L}{2\pi \xi_x(L)} \, .
\label{Xxi}
\end{equation}
For a system at criticality, this relation may be expected to hold
only in the limit of $L \to \infty$, because the irrelevant fields
that are usually present cause deviations from the conformal symmetry.
But, in many cases, the calculation of $\xi$ for a limited number of
system sizes still allows a reasonably accurate estimate of the
scaling dimension $X$. The correlation length $\xi_x(L)$ is related
to an eigenvalues $\Lambda_x$ of the transfer matrix as
\begin{equation}
 \xi_x^{-1}(L) = \zeta \ln \frac{\Lambda_0(L)}{\Lambda_x(L)}
\label{xilambda}
\end{equation}
where where the geometrical factor $\zeta$ (the ratio between the unit
of $L$ and the thickness of a layer added by $\mat{T}$), $\Lambda_0(L)$ 
is the largest eigenvalue of $\mat{T}$, and $\Lambda_x(L)$ the
eigenvalue associated with the connected correlation function between
the two vertices. One of the remaining tasks is thus to identify the
latter eigenvalue in the spectrum of $\mat{T}$.

\subsection{Block structure of the transfer matrix}
To determine the eigenvalue $\Lambda_x(L)$,
we divide the connectivities in two groups: the first group contains
the well-nested ones labeled by a subscript ${\rm w}$, and the second 
group the entangled ones,  labeled by a subscript ${\rm e}$.
The transfer matrix can then, in obvious notation, be divided in four
blocks as
\begin{equation}
\mat{T}= \left[
\begin{array}{rr}
 \mat{T}_{\rm ww}  & \mat{T}_{\rm we} \\
 \mat{0}_{\rm ew}  & \mat{T}_{\rm ee}
\end{array}
\right] \, ,
\end{equation}
where the lower left block contains only zeroes, because the
Hamiltonian does not contain crossing bonds, and the transfer matrix 
is thus unable to form entangled states out of well-nested ones.
As a consequence, the {\em transpose} transfer matrix $\mat{T}^{\rm t}$
cannot form well-nested states out of entangled ones. The eigenvalue
problem of $\mat{T}$ in effect decomposes in the two separate
eigenvalue problems of $\mat{T}_{\rm ww}$ and $\mat{T}_{\rm ee}$.
Our algorithm
to find the eigenvalues \cite{BN1982} is based on the analysis of a 
sequence of vectors obtained by the repeated multiplication of an
initial vector by $\mat{T}^{\rm t}$. Thus, if we use an initial
vector that contains only entangled states, we easily obtain the
largest eigenvalue of the diagonal block $\mat{T}_{\rm ee}$. 
Naturally, the largest eigenvalue of $\mat{T}_{\rm ee}$ will be
associated  with the `least entangled states' in which only two
clusters are entangled, as for instance the four-site connectivity
in which site 1 is connected to 3, and 2 to 4. That is precisely
the set of connectivities describing the correlation between two of
the aforementioned vertices, which, in the geometry of an infinitely
long cylinder, are thought to be located at $\pm \infty$.

\subsection{Finite-size analysis}
The eigenvalues $\Lambda_0(L)$ and $\Lambda_x(L)$ were calculated
numerically as described above, for finite sizes up to $L=15$, for
which the number of connectivities is 390~248~055. This was done
for two different transfer matrices. First, we chose the transfer
direction parallel to a set of lattice edges, and second, we chose
it in the direction of a set of diagonals of the elementary faces of
the square lattice. The latter method can handle larger system sizes
when expressed in lattice edges, although the number of finite
sizes is the same. Since the finite-size parameter $L$ denotes the
number of sites added by a multiplication by $\mat{T}^{\rm t}$,
the circumference of the cylinder is $L$ lattice edges for the
`parallel' transfer matrix, and $L \sqrt 2$ lattice edges for the
`diagonal' transfer matrix. The geometric factors are $\zeta=1$ and
$\zeta=1/2$ respectively, because we express $\xi_x(L)$ in the same
units as $L$.

\begin{figure}
\includegraphics[scale=0.9]{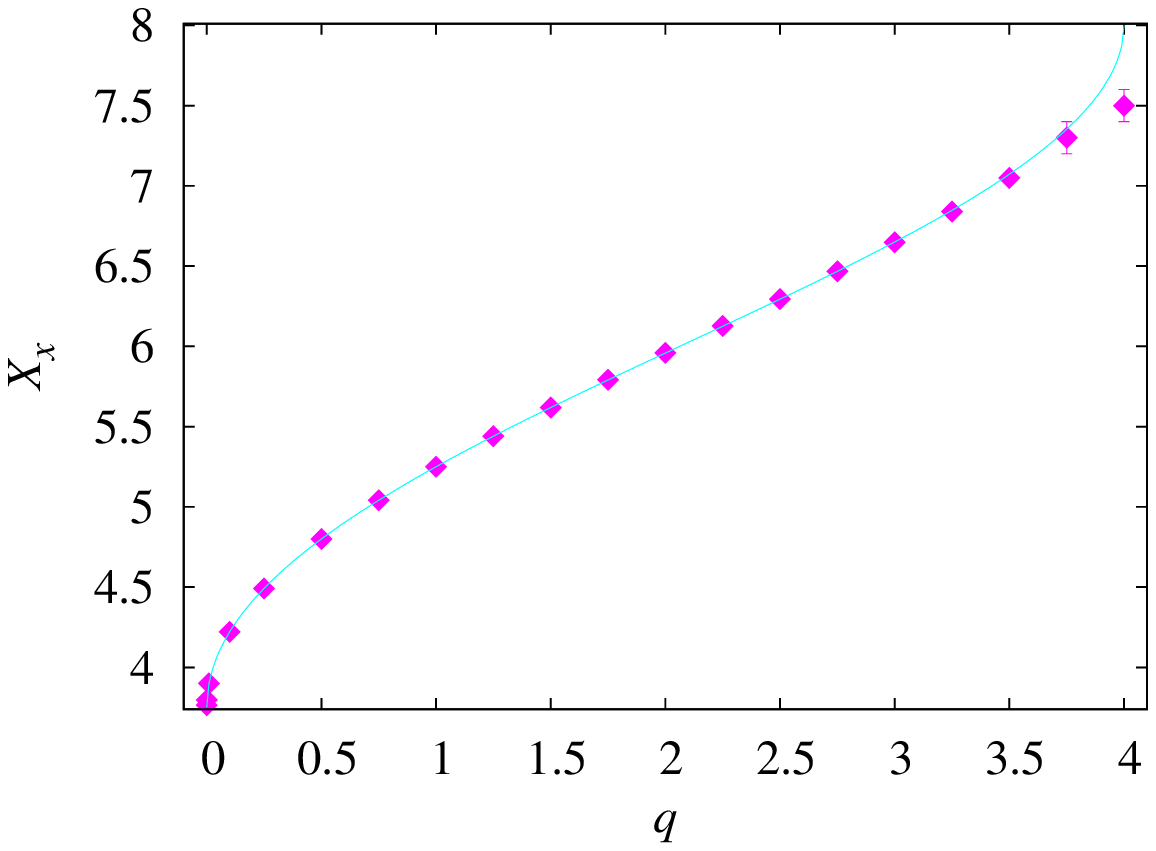}
\caption{Results for the scaling dimension $X_x$ associated with crossing
bonds in the random-cluster model, as a function of the number $q$ of
Potts states. 
The symbols show the averages of the numerical results obtained from the
two transfer matrices. The curve shows the Coulomb gas result.}
\label{figx1}
\end{figure}

A convenient quantity in the finite-size analysis is the 'scaled
gap' which depends here (except on the type of the transfer matrix)
only on $L$, because the Potts
coupling is set at its critical value. This quantity is defined as
\begin{equation}
X_x(L)= L/[2\pi \xi_x(L)] \, .
\end{equation}
In the vicinity of a renormalization fixed point, finite-size
scaling leads to the equation
\begin{equation}
X_x(L) = X_x + a_u L^{y_u} + \ldots
\label{X-scaling}
\end{equation}
where $y_u$ is an irrelevant exponent and $a_u$ the associated
finite-size amplitude, and the dots stand for further finite-size
corrections. The first step of the estimation of the scaling dimension
$X_x$ is done by means of power-law fits according to this equation.
The resulting estimates are expected to depend, again, on $L$ as a
power law.  Thus, better estimates can be obtained by means of an
iterated fitting procedure. Up to four iteration steps were made.
We tried several variations of the fitting procedure, concerning
the use of a-priori knowledge of the exponents of the finite-size
dependences. We expect corrections described by the integer exponent
$-2$, as well as by the irrelevant Potts exponent \cite{BN2}.
Further details appear in Refs.~\onlinecite{BN1982} and \onlinecite{FSS}. 
The numerical uncertainties in $X_x$ were roughly estimated from the
variation of its fitted value with increasing system size.
The best estimates, and the estimated error margins, are shown in
Table  \ref{tab_1} and Fig.~\ref{figx1}, together with the
theoretical values derived in Sec.~\ref{CG}.

\begin{table}
\caption{Results of transfer-matrix calculations for the crossing-bond
dimension. The numerical results for $X_x$ are indicated by the 
superscript `num'. They were determined using two
different transfer-matrix methods: with the transfer direction 
parallel to a set of edges of the square lattice, as indicated by
`(e)', and with transfer direction parallel to a set of diagonals 
of the elementary faces, as indicated by `(d)'. The estimated error
in the last decimal place is shown between parentheses. For comparison
we also show the Coulomb gas prediction in the last column. }
\label{tab_1}
\begin{center}
\begin{tabular}{||l|lr|lr|c||}
\hline
$q$  &$X_x^{\rm num}$(e)& error&$X_x^{\rm num}$(d)& error&$X_x{\rm theory}$\\
\hline
0.0001 &     3.76511     &  (1) &   3.765110      & (2)  &     3.765110    \\
0.001  &     3.79772     &  (1) &   3.797715      & (2)  &     3.797714    \\
0.01   &     3.90028     &  (1) &   3.900277      & (2)  &     3.900278    \\
0.10   &     4.22087     &  (1) &   4.220862      & (5)  &     4.220863    \\
0.25   &     4.49180     &  (2) &   4.49180       & (1)  &     4.491800    \\
0.50   &     4.7997      &  (1) &   4.79970       & (5)  &     4.799728    \\
0.75   &     5.0411      &  (1) &   5.0410        & (1)  &     5.040971    \\
1.00   &     5.2500      &  (1) &   5.2500        & (1)  &     5.250000    \\
1.25   &     5.4404      &  (1) &   5.4404        & (1)  &     5.440283    \\
1.50   &     5.6190      &  (2) &   5.6192        & (2)  &     5.618945    \\
1.75   &     5.790       &  (1) &   5.7912        & (5)  &     5.790520    \\
2.00   &     5.957       &  (1) &   5.959         & (1)  &     5.958333    \\
2.25   &     6.123       &  (2) &   6.127         & (2)  &     6.125203    \\
2.50   &     6.290       &  (5) &   6.295         & (2)  &     6.293876    \\
2.75   &     6.46        &  (1) &   6.468         & (2)  &     6.467453    \\
3.00   &     6.63        &  (2) &   6.648         & (5)  &     6.650000    \\
3.25   &     6.80        &  (5) &   6.84          & (1)  &     6.847755    \\
3.50   &     7.1         &  (1) &   7.05          & (1)  &     7.072311    \\
3.75   &     7.3         &  (2) &   7.3           & (1)  &     7.353038    \\
4.00   &     7.4         &  (5) &   7.5           & (1)  &     8.000000    \\
\hline
\end{tabular}

\end{center}
\end{table}

\section {Discussion }\label{Dis}
The introduction of next-nearest-neighbor couplings in Potts models
increases the size of the transfer matrix, and thus leads to new
eigenvalues of which we have associated one with the scaling dimension
$X_x$. Our analysis was only focused on the determination of $X_x$.
Of course, the appearance of effects described by this scaling dimension
is only one of the consequences. The new couplings do also increase
the range of the interactions, and this influences the irrelevant field
and its associated corrections to scaling \cite{Wegner}. This effect
is well known and outside the scope of the present paper.

Most of the numerical data in Table \ref{tab_1} agree well with the
Coulomb gas prediction. But it is clear that the agreement deteriorates
near $q=4$, where the error estimates become quite large, but not large
enough to explain the difference with the Coulomb gas result. This
problem is explained in terms of the behavior of the second thermal
exponent of the Potts model, which is strongly irrelevant for small $q$,
but increases with $q$ and becomes marginal at $q=4$. This explains why
the fast convergence at small $q$ deteriorates for larger $q$, in
particular at $q=4$ where a logarithmic correction factor leads to
misleading finite-size fits, which assume a power-law behavior instead.
We thus conclude that our numerical analysis is in agreement with the
theoretical predictions.

The consistency between the theory and the numerical results for the
scaling dimension associated with crossing bonds is quite reassuring.
However, a strange aspect of the present result is that it seems 
inconsistent with the equivalence between the O(1) model and the
$q=2$ Potts or random-cluster model. Both of these are Ising models,
and one may thus expect to find the same scaling dimension for crossing 
bonds. For the critical O(1) model, one finds $X_x=21/8$ \cite{BN2},
whereas Eq.~(\ref{Xx}) yields $X_x=143/24$. To answer this paradox
one may note that, for the Ising case, the scaling behavior of the
thermodynamic observables is determined by a very limited set of
scaling fields associated with primary operators in the conformal
field theory \cite{JLC1}. The critical amplitudes associated with
other scaling fields vanish in thermodynamic quantities but may
appear in e.g. the fractal dimensions of spin clusters or percolation
clusters defined on the ensemble of configurations generated by
Eqs.~(\ref{Potts}) and (\ref{Zrc}). The geometric correlations
associated with crossing random-cluster bonds should naturally be
different from those associated with crossing bonds in the O($1$)
model because the two corresponding Coulomb gas descriptions are
based on two different graph expansions of the Ising model.

In the notation based on the Kac formula \cite{FQS} for the dimensions
of rotationally invariant observables, namely
\begin{equation}
X_{p,q} =\frac{[p(m+1)-qm]^2-1}{2m(m+1)} \, .
\end{equation}
where $m=g/(1-g)$ for the critical Potts model,
the Coulomb gas result Eq.~(\ref{Xx}) can be written
\begin{equation}
X_x= X_{0,4} \, .
\end{equation}
This is outside the restricted set of operators \cite{JLC1} with
$1\leq p < m$, $1\leq q < m+1$ which describe the thermodynamics 
of unitary models. Thus, for critical Potts models with integer $q$,
effects described by the dimension $X_x$ should be absent in the
thermodynamic behavior, but may still appear in fractal and geometric
properties of critical spin configurations.

Finally, we note that the Coulomb gas result Eq.~(\ref{Xx}) for
the dimension $X_x$  also applies to the tricritical branch of
the random-cluster model, with $g=1+ {\rm arccos}(q/2-1)/(2 \pi)$ 
instead of  Eq.~(\ref{gdef}), which implies that the crossing
bond dimension is even larger, i.e., more irrelevant, than that on
the critical branch. We recall that, in contrast, the crossing-bond
exponent is {\em relevant} in the tricritical O($n$) model.

\acknowledgments
We are indebted to Prof. B. Nienhuis for valuable discussions.
H.B. gratefully acknowledges the hospitality of the Physics Department
of the Beijing Normal University and the Modern Physics Department of
University of Science and Technology of China. The research is supported
by the Science Foundation of The Chinese Academy of Sciences, by the NSFC 
under Grant \#10675021, by the Program for New Century Excellent 
Talents in University (NCET), and by the Lorentz Fund.

\end{document}